\documentclass[11pt]{article}
\usepackage[dvipdfmx]{graphicx}
\usepackage{color}
\usepackage{amsmath,amssymb,amsfonts,latexsym}
\begin{document}

\begin{center}
\textbf{Emperical Study on Various Symmetric Distributions for Modeling Time Series}

\vspace{8mm}
{\large Genshiro Kitagawa}\\[2mm]
Tokyo University of Marine Science and Technology\\[-1mm]
and\\[-1mm]
The Institute of Statistical Mathematics

\vspace{3mm}
{\today}
\end{center}


\noindent
\textbf{Abstract:}

This study evaluated probability distributions for modeling time series with abrupt structural changes. The Pearson type VII distribution, with an adjustable shape parameter $b$, proved versatile. The generalized Laplace distribution performed similarly to the Pearson model, occasionally surpassing it in terms of likelihood and AIC. Mixture models, including the mixture of $\delta$-function and Gaussian distribution, showed potential but were less stable. Pearson type VII and extended Laplace models were deemed more reliable for general cases. Model selection depends on data characteristics and goals.

\vspace{2mm}
\noindent
\textbf{Key words:}
Non-Gaussian state-space model; non-Gaussian filter/smoother; trend estimation; Pearson type VII distribution; genralized Laplace distribution; mixture distributions.

\section{Introduction}

The state-space model is a highly effective time-series model representation framework that can accommodate numerous time-series models in a unified manner. Once a linear Gaussian state-space representation is obtained, the state vector can be efficiently estimated sequentially from the observed data using a Kalman filter and smoother. This property enables the unified estimation of parameters, prediction, smoothing, decomposition into several components, and interpolation of missing values for time series models.
Indeed, state-space models are employed in various fields, including seasonal adjustment methods, long-term prediction, and system control (Harison and Steevens (1976),  West and Harrison (1989), Kitagawa and Gersch (1996), Shumway and Stoffer (2000), Durbin and Koopman (2012).

Although linear Gaussian state space models have considerable advantages, they are not universal. For example, sudden changes or bending of the trend, structural changes in the system or the presence of outliers do not give good results with linear Gaussian state-space models, but non-Gaussian probability distributions can sometimes solve these problems.

In this paper, we take artificial data with structural changes as an example and consider the characteristics of various probability distributions in terms of influence functions and trend posterior distributions.
Although the example taken is a very simple one, the findings on the characteristics of non-Gaussian distributions obtained here should be useful for more complex models.

The paper is structured as follows.
Section 2 briefly summarises the state-space model and the problems of state and parameter estimation, and presents the results of trend estimation using an ordinary Gaussian distribution for the data used below.
Section 3 briefly introduces the state estimation methods for the case of non-Gaussian distributions, presents the density and influence functions for the three cases of Pearson family of distributions, generalized Laplace distribution and mixture of some probability distributions, and compares the trend estimation results using these noise distributions.
Section 4 presents a summary of the empirical facts obtained in Section 3.

\section{A Brief Review of the Filtering and Smoothing Algorithms}

\subsection{The state-space model and the state estimation problems}

Assume that a time series \( y_n \) is expressed by a linear state-space model 
\begin{eqnarray}
x_{n} &=& F_nx_{n-1} \: + \: G_{n}v_{n} \label{ssm-1} \nonumber \\
y_n &=& H_nx_n \: + \: w_n, \label{ssm-1}
\end{eqnarray}
where \(x_n\) is an \( k \)-dimensional state vector,
\( v_{n} \) and \( w_n \) are \( \ell \)-dimensional and 1-dimensional white noise sequences having density functions \(q_{n}(v) \) and \( r_n(w) \), respectively.
The initial state vector \( x_0 \) is assumed to be distributed
according to the density \( p(x_0 ) . \)

The information from the observations up to time \( j \) is denoted by \(Y_j \),
namely, \( Y_j \equiv \{y_1,\ldots ,y_j\}\).
The problem of state estimation is to evaluate \( p(x_n|Y_j)\), 
the conditional density of \( x_n \)
given the observations \(  Y_j \) and the initial density \( p(x_0|Y_0) \equiv p(x_0) . \)
For \(n>j, n=j \) and \(n<j \), it is called the problem of prediction, filtering
and smoothing, respectively.

It is well-known that the conditional density of $x_n$ given $Y_n$, \( p(x_n|Y_m) \), is also Gaussian and that the mean and the variance covariance matrix
can be obtained by the Kalman filter and the fixed interval smoothing algorithms (Anderson and Moore, 1979).

\subsection{Parameter Estimation for the State-Space Model}

State-space models usually include several unknown parameters such as the system noise variance $\tau^2$ and the  observation noise variance $\sigma^2$, etc. These are collectively expressed as a parameter vector $\theta$.
Given the one-step-ahead prediction of the state vector, $x_{n|n-1}$ and its variance and covariance matrix, $V_{n|n-1}$, the log-likelihood of the state-space model is given by
\begin{eqnarray}
\ell (\theta) &=& \log L(\theta ) \nonumber \\
  &=& -\frac{1}{2}\sum_{n=1}^N \log (2\pi r_n)
      -\frac{1}{2}\sum_{n=1}^N \frac{\varepsilon_n^2}{2r_n},
\end{eqnarray}
where $\varepsilon_n = y_n - Hx_{n|n-1}$ and $r_n = \sigma^2_n + H_n V_{n|n-1} H_n^T$.
Therefore, the maximum likelihood estimate of the parameter $\theta$ is obtained by maximizing this log-likelihood through numerical optimization.

\subsection{Test Example and State-Space Model for Trend Estimation}

\begin{figure}[h]
\begin{center}
\includegraphics[width=75mm,angle=0,clip=]{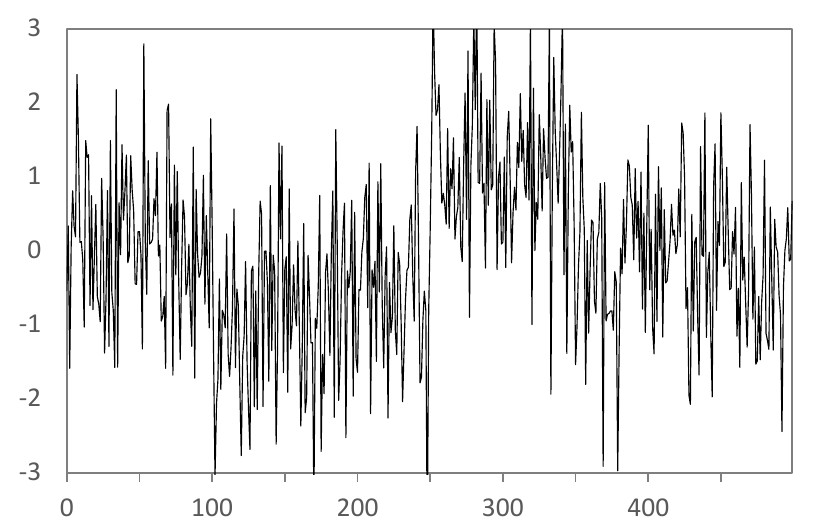}
\includegraphics[width=75mm,angle=0,clip=]{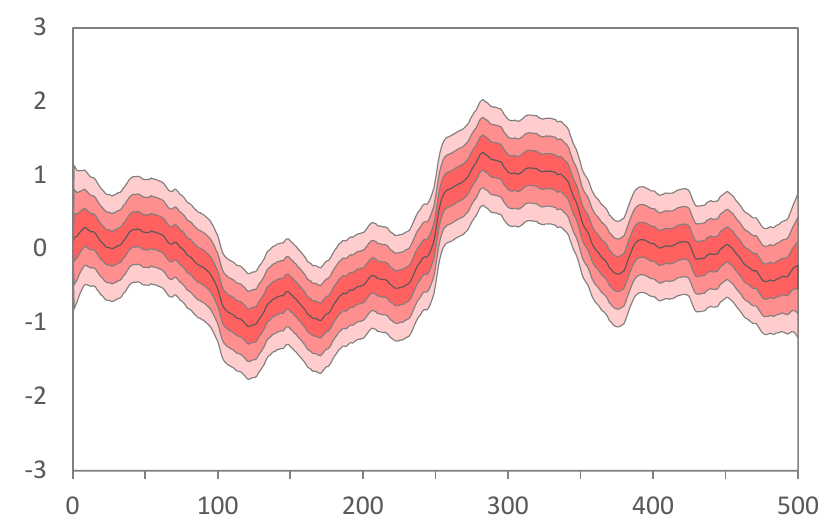}
\caption{Test data and trend and the marginal posterior distribution obtained by Gaussian noise model.
The posterior distribution is expressed by the posterior mean (bold curve) and $\pm$1, 2 and 3
standard errors.}
\label{figure:test_data_and_Gauss}
\end{center}
\end{figure}

The left panel of the Figure \ref{figure:test_data_and_Gauss} shows the test data considered in this paper. The number of data, $N$, is 500 and at least three jumps can be seen in the middle of the data (Kitagawa 1987).
To estimate the trend, we consider the following first-order trend model, i.e. the random walk model.
\begin{eqnarray}
\begin{array}{cclcl}
t_n &=& t_{n-1} + v_n  \\
y_n &=& t_n + w_n,    
\end{array}   \label{Eq_state-space_model}
\end{eqnarray}
where $y_n$ is the observed time series, $t_n$ is unobservable trend component, $v_n$ is the system noise and $w_n$ is the observation noise.
It is assumed that $v_n$ and $w_n$ are white noise sequence following
$v_n \sim N(0,\tau^2)$ and $w_n \sim N(0,\sigma^2)$.
By putting $F=G=H=[1]$, it can be seen that this model itself is the simplest state-space model.

The right panel of Figure \ref{figure:test_data_and_Gauss} shows the trend estimates obtained by the usual Kalman filter and smoother.
The maximum likelihood estimate of the parameters are $\hat{\tau}^2=1.022\times 10^{-2}$ and 
$\hat{\sigma}^2=1.043$.
The thick line in the middle shows the trend estimate and the three lines above and below it respectively show the $\pm$1, 2 and 3 standard errors.
The estimated trend undulates in response to fluctuations in the data. However, not only are abrupt changes not fully captured, but also the fluctuations of the trend during the period appear to be too large. 

To address this problem, the following sections assume various non-Gaussian distributions for the system noise $v_n$ and compare the results. Although only the simplest case of trend estimation is presented in this paper, it is believed that it can also be used in many state-space modelling applications.

\section{Non-Gaussian Noise Distribution Models}

We assume the same linear state-space model as in equation (\ref{Eq_state-space_model}). However, unlike in (\ref{Eq_state-space_model}), the system noise $v_n$ is assumed to be a general probability distribution following a density function $q(x)$.

Needless to say, if the noise distribution is not Gaussian, as in this case, there is no guarantee that the Kalman filter will yield good estimates. Generally, nonlinear filters such as particle filters are used (Gordon et al., 1993; Kitagawa 1996), but as the example considered here is a simple case with a one-dimensional state vector, we will use non-Gaussian filters and smoothing, which can be computed more rigorously. However, the various noise distributions considered here can also be used in particle filter computations.

\subsection{The non-Gaussian filter and the smoother}

 It is well-known that for the nonlinear and/or non-Gaussian state-space model, 
the recursive formulas for obtaining the densities of the one step ahead predictor, the filter and the smoother are as follows:\\
{\bf One step ahead prediction:}
\begin{equation}
 p(x_{n} |Y_{n-1} )
  = \int_{- \infty}^\infty p(x_{n} | x_{n-1} )
p(x_{n-1} |Y_{n-1} )dx_{n-1}. 
\end{equation}
{\bf Filtering:}
\begin{equation}
 p(x_n |Y_n ) = \frac{p(y_n|x_n)p(x_n|Y_{n-1})}{\int p(y_n |x_n )p(x_n |Y_{n-1} )dx_n}.
\end{equation}
{\bf Smoothing:}
\begin{equation}
 p(x_n |Y_N ) 
  = p(x_n |Y_n ) \int_{- \infty}^\infty \frac{p(x_{n+1} |Y_N )
p(x_{n+1} |x_n )}{p(x_{n+1} |Y_n )}dx_{n+1} .
\end{equation}

 In Kitagawa (1987, 1988), an algorithm for implementing the non-Gaussian filter and smoother was developed by approximating each density function using a step-function or a continuous piecewise linear function  and by performing numerical computations.
This method was successfully applied to various problems 
(Kitagawa, 1987, 2020; Kitagawa and Gersch, 1996).

The following subsections compare and contrast the differences in terms of the influence function (Hampel et al. 1986) and the posterior distribution of the trend for the three cases of the Pearson type VII family of distributions, the generalized Laplace distribution and various mixture of two distributions.

\subsection{Pearson Type VII Distributions}

The density function of the Pearson type VII family of distributions is given by
\begin{eqnarray}
p_b(x) = \frac{C}{(x^2+\tau^2)^b} ,
\end{eqnarray}
where $\frac{1}{2} < b < \infty$ is the shape parameter, $\tau^2$ is the dispersion parameter and $C = \tau^{2b-1}\Gamma (b)/\Gamma(\frac{1}{2})$ $\Gamma(b-\frac{1}{2})$.

This family of Pearson distributions contains a wide range of heavily-tailed symmetric probability distributions, with a Cauchy distribution when $b$=1 and a $t$-distribution with $k$ degrees of freedom when $b=(k+1)/2$. It is also a normal distribution when $b\rightarrow\infty$.
The influence function of the Pearson family of distributions is given by
\begin{eqnarray}
-\frac{d\log p_b(x)}{dx} = \frac{2bx}{x^2 + \tau^2}.
\end{eqnarray}

\begin{figure}[tbp]
\begin{center}
\includegraphics[width=150mm,angle=0,clip=]{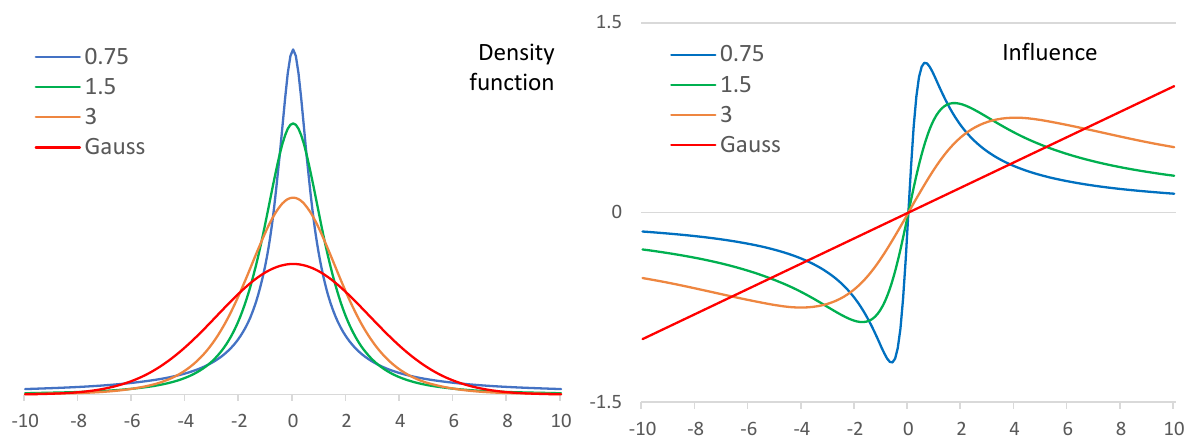}
\caption{Density functions and influence functions of the Pearson type VII family of distributions with with $b$=0.75, 1.5, 3.}
\label{figure:Influence_Pearson}
\end{center}
\end{figure}

As shown in the left-hand side of Figure \ref{figure:Influence_Pearson}, the density function of the Pearson distribution family becomes more centred and sharper as the value of $b$ decreases.
As shown by the red line in the right-hand plot, the influence function of the Gaussian distribution is linear, but as the value of $b$ decreases, the influence near the origin increases and the influence away from the origin decreases.

Table \ref{Tab_accuracy_of posterior_mean} shows the results of estimating the trend of the test data in Figure \ref{figure:test_data_and_Gauss} using the Pearson distribution for various $b$-values for system noise, showing the observed noise variance $\sigma^2$, system noise variance $\tau^2$, log-likelihood, the number of free parameters and AIC obtained by the maximum likelihood estimation method.

\begin{table}[tbp]
\caption{Trend model with Pearson type VII noise distributions with various $b$.}
\label{Tab_accuracy_of posterior_mean}
\tabcolsep=2mm
\begin{center}\begin{tabular}{ccccccc}
 $b$  & $\sigma^2$ & $\tau^2$ & log-LK & $k$ & AIC \\
\hline
0.55  & 1.0275 & $1.930\times 10^{-13}$ & $-742.6529$ &  2 & 1,489.306   \\
0.60  & 1.0264 & $2.613\times 10^{-12}$ & $-742.3849$ &  2 & 1,488.770   \\
0.75  & 1.0227 & $1.694\times 10^{-8}$  & $-741.8962$ &  2 & {\color{red}1,487.792}   \\
1.00  & 1.0227 & $3.454\times 10^{-5}$  & $-742.1673$ &  2 & 1,488.335   \\
1.50  & 1.0198 & $2.602\times 10^{-3}$  & $-743.7025$ &  2 & 1,491.405   \\  
2.00  & 1.0189 & $1.072\times 10^{-2}$  & $-745.1989$ &  2 & 1,490.398   \\
3.00  & 1.0265 & $3.667\times 10^{-2}$  & $-747.3061$ &  2 & 1,498.612   \\
$\infty$&1.0439& $1.249\times 10^{-2}$  & $-748.5156$ &  2 & 1,501.031   \\
\hline
\end{tabular}\end{center}
\end{table}

\begin{figure}[tbp]
\begin{center}
\includegraphics[width=75mm,angle=0,clip=]{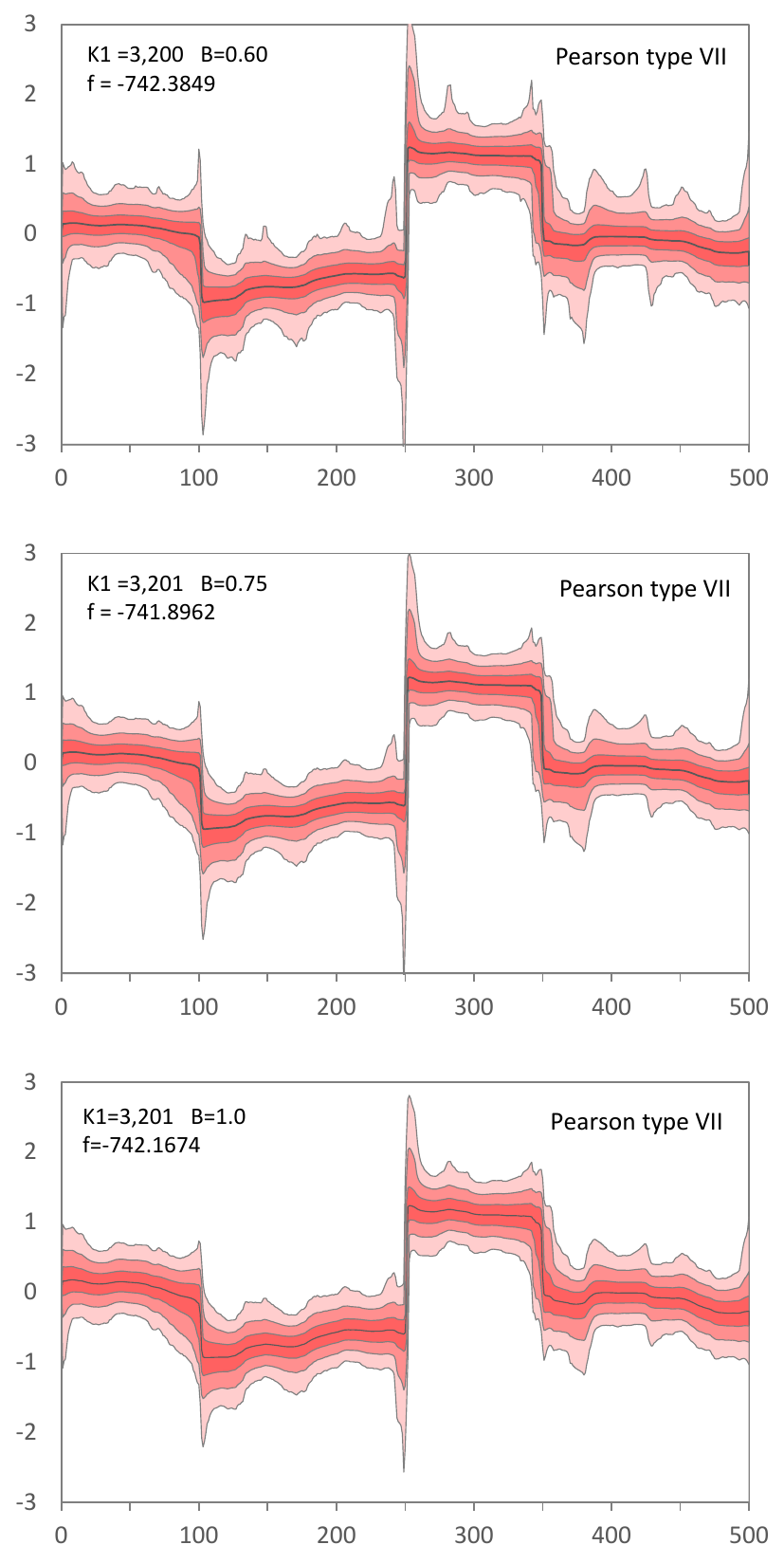}
\includegraphics[width=75mm,angle=0,clip=]{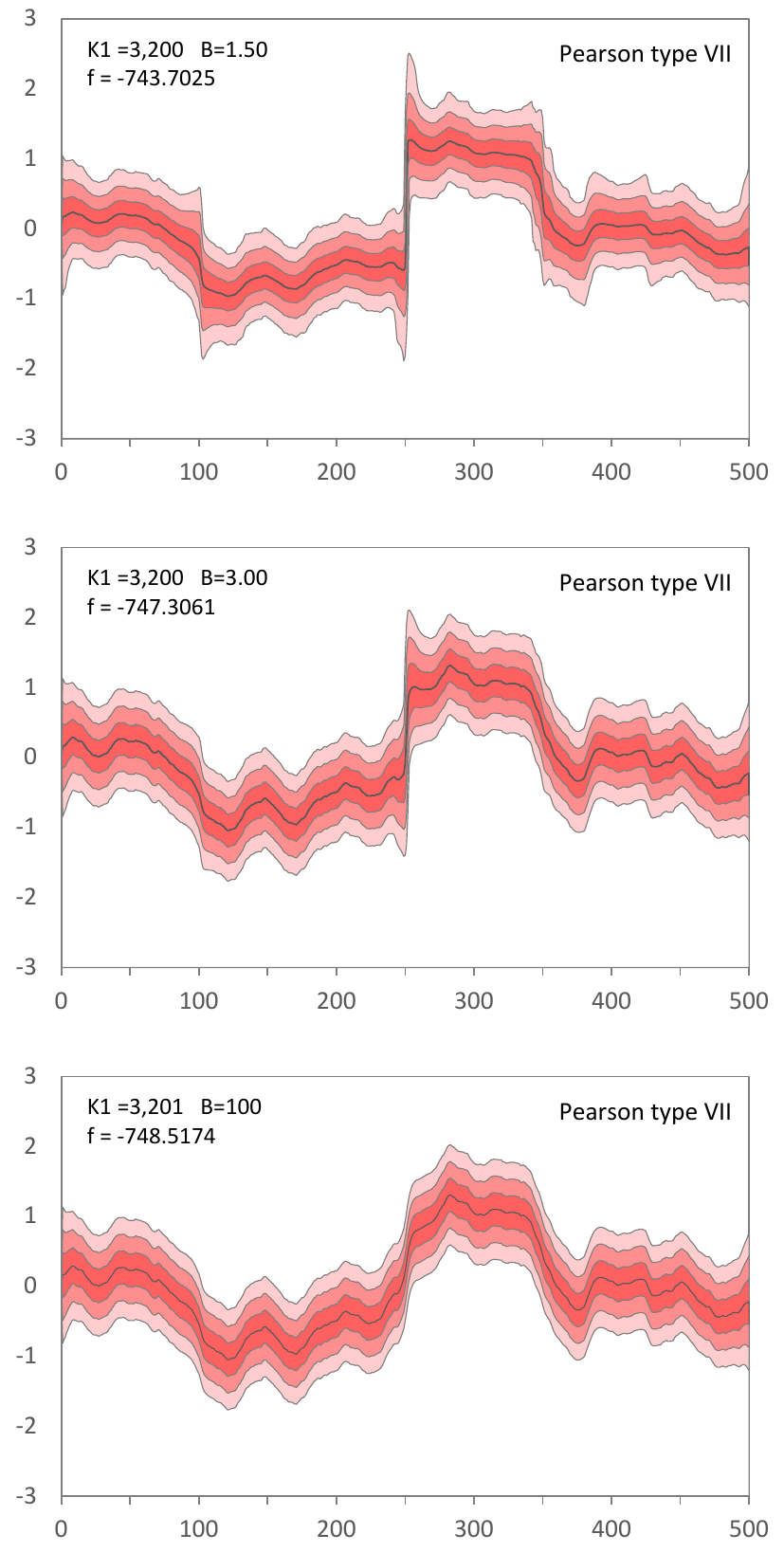}
\caption{Trend estimation by self-organizing state-space model implemented by non-Gaussian filter and smoother.
From upper left to lower right, Pearson type VII distributions with $b$=0.60, 0.75, 1.0, 1.50, 3.0 and 100.}
\label{figure:trend_Pearson}
\end{center}
\end{figure}

The Gaussian distribution ($b=\infty$) has the largest AIC and is shown to be the worst fit in the Pearson family of  distributions when there are jumps, as in the test data. The AIC value is smallest at $b$=0.75, and this distribution is sharper than the Cauchy distribution.
However, the change in AIC is not significant below $b$=1.

Figure \ref{figure:trend_Pearson} shows the posterior distribution of the trend for the Pearson distributions of several $b$ values as system noise; the seven curves show the 0.13, 2.27, 15.87, 50.5, 84.13, 97.73 and 99.87 percentile points, corresponding to the mean and $\pm$1, 2 and 3 standard deviations of the Gaussian distribution.
In the bottom right plot, $b$=100 is for a Gaussian distribution, and this result is identical to the right plot of Figure \ref{figure:test_data_and_Gauss}. 
No jumps are detected and an undulating trend is obtained. 
For $b$=3.0 in the middle right plot, a jump in the trend is detected only around $n$=250 in the middle.
The three figures on the left with $b$-values below 1 show almost identical results, with jumps at $n$ = 100, 250 and 350 being detected successfully, and a very smooth trend estimated elsewhere.

\subsection{Generalized Laplace Distributions}

The density function of the generalized Laplace distributions is given by (Kotz et al. 2001)
\begin{eqnarray}
p_b(x) = C \exp\{ -\tau |x|^b \}
\end{eqnarray}
where $\frac{1}{2} < b < \infty$ is the shape parameter, $\tau^2$ is the dispersion parameter and $C = \tau^{2b-1}\Gamma (b)/\Gamma(\frac{1}{2})$ $\Gamma(b-\frac{1}{2})$.

The Generalized Laplace distribution becomes a Laplace distribution (two-sided exponential distribution) when $b$=1 and a Gaussian distribution when $b$=2, but for modelling purposes it is often used when $b$ is less than 1. In such cases, the density function becomes a very sharp distribution at the origin.
The influence function of the Pearson family of distributions is given by
\begin{eqnarray}
-\frac{d\log p_b(x)}{dx} = \left\{  \begin{array}{cc} \tau b|x|^{b-1}  & x > 0 \\
                                                     -\tau b|x|^{b-1}  & x<0
                                    \end{array} \right.
\end{eqnarray}

The left plot in Figure \ref{figure:Influence_Generalized Laplace} shows the density function of the generalized Laplace distribution and the right plot shows the influence function: when $b$ is less than 1, the density function is not differentiable at the origin $x$=0 and the influence function cannot be defined; when $b \rightarrow$0, the density function is asymptotic to the mixture of the delta function and the uniform distribution, as shown later. The influence function becomes a step function when $b$=1, as shown by the red line, but for $b>1$ it is weighted even more extremely than the Pearson family of distributions.

\begin{figure}[h]
\begin{center}
\includegraphics[width=150mm,angle=0,clip=]{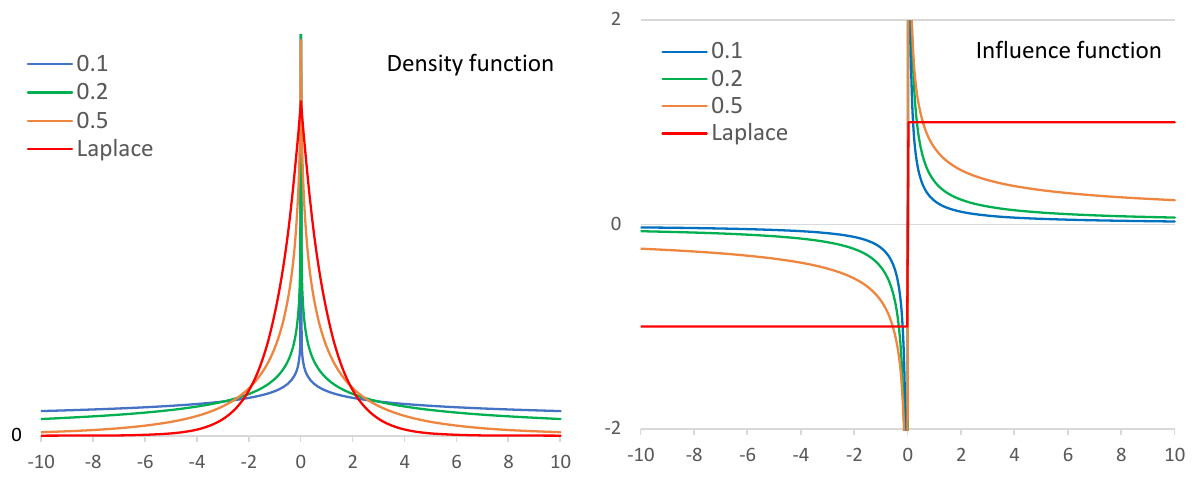}
\caption{Density functions and influence functions of the generalized Laplace distributions with $b$=0.1, 0.2, 0.5 and 1.}
\label{figure:Influence_G-Laplace}
\end{center}
\end{figure}

Table \ref{Tab_summary_of_G-Laplace_distribution_models} shows the results of estimating the trend of the test data using the Generalized Laplace distribution for various $b$-values for system noise, showing the observation noise variance $\sigma^2$, system noise variance $\tau^2$, log-likelihood, the number of free parameters and AIC obtained by the maximum likelihood estimation method.

The AIC is minimum at $b$=0.01, yielding a better model than the Laplace distribution model ($b$=1) or the Gaussian distribution model ($b$=2).
\begin{table}[tbp]
\caption{Trend models with generalized Laplace noise distributions with various $b$.}
\label{Tab_summary_of_G-Laplace_distribution_models}
\tabcolsep=2mm
\begin{center}\begin{tabular}{ccccccc}
 $b$  & $\sigma^2$ & $\tau^2$ & log-LK & $k$ & AIC \\
\hline
0.05  & 1.0298 & 16.275 & $-742.1611$ &  2 & 1,488.322   \\
0.10  & 1.0251 & 15.928 & $-741.7902$ &  2 & {\color{red}1,487.580}   \\
0.20  & 1.0198 & 13.629 & $-742.8129$ &  2 & 1,489.626   \\
0.50  & 1.0181 &\,9.869 & $-745.6362$ &  2 & 1,495.272   \\
1.00  & 1.0359 & 12.335 & $-747.9692$ &  2 & 1,499.938   \\  
2.00  & 1.0439 & 40.031 & $-748.5156$ &  2 & 1,501.031   \\
4.00  & 1.0459 & 754.43 & $-748.6541$ &  2 & 1,501.308 \\
\hline
\end{tabular}\end{center}
\end{table}

\begin{figure}[tbp]
\begin{center}
\includegraphics[width=75mm,angle=0,clip=]{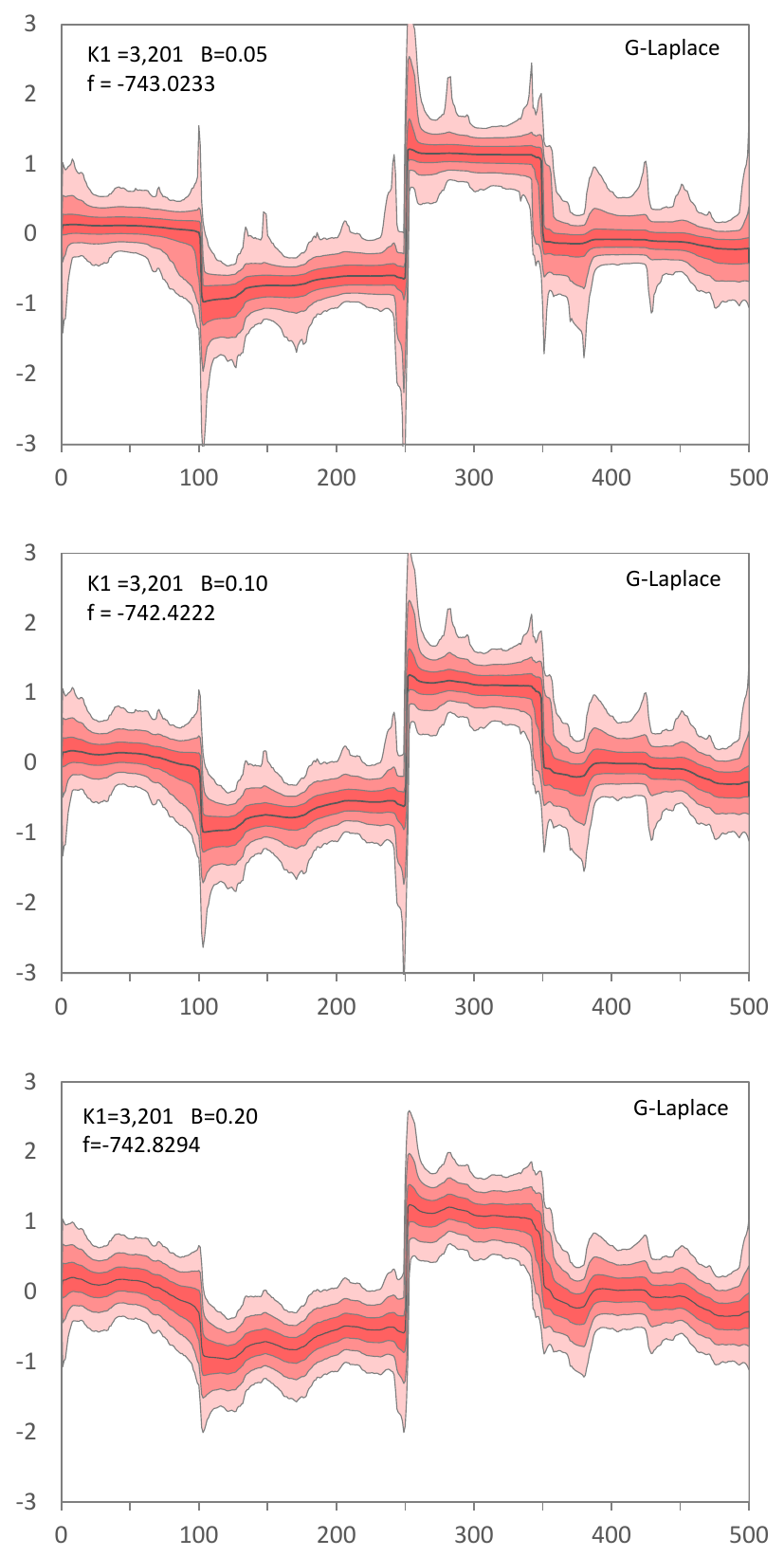}
\includegraphics[width=75mm,angle=0,clip=]{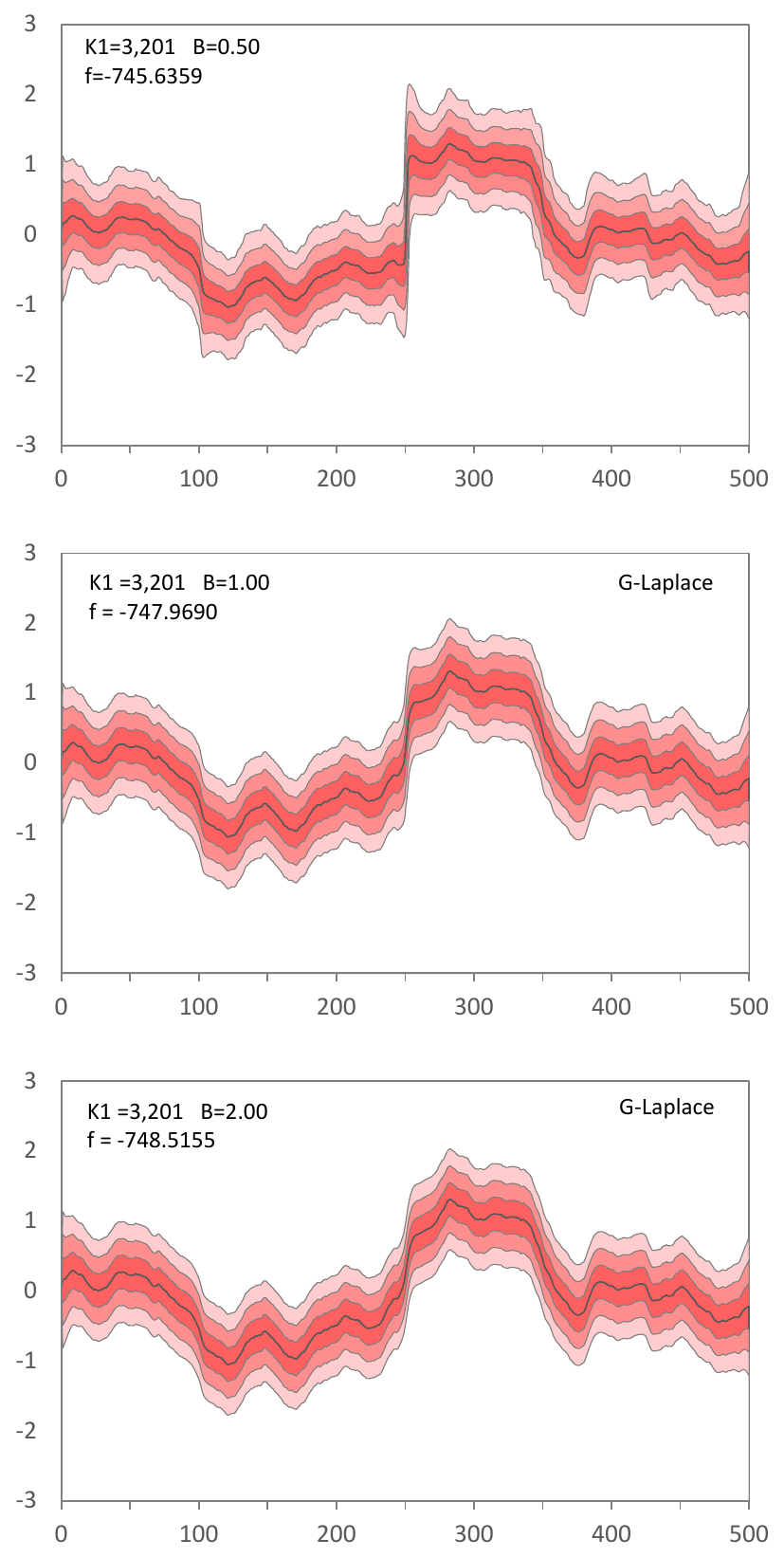}
\caption{Trend estimation by self-organizing state-space model implemented by non-Gaussian filter and smoother.
From upper left to lower right, Generalized Laplace distributions with $b$=0.05, 0.10, 0.20, 0.50, 1.0 and 2.0.}
\label{figure:trend_G-Laplace}
\end{center}
\end{figure}

Figure \ref{figure:trend_G-Laplace} shows the posterior distribution of the trend for the generalized Laplace distribution of several $b$ values as system noise.
In the bottom right plot, $b$=100 is for a Gaussian distribution, and this result is identical to the right plot of Figure \ref{figure:test_data_and_Gauss}. 
No jumps are detected and an undulating trend is obtained. 
For $b$=3.0 in the middle right plot, a jump in the trend is detected only around $n$=250 in the middle.
The three figures on the left with $b$-values below 1 show almost identical results, with jumps at $n$ = 100, 250 and 350 being detected successfully, and a very smooth trend estimated elsewhere.

The bottom right figure shows the case of the Gaussian distribution ($b$=2) and the middle right figure shows the Laplace distribution ($b$=1); the AIC is smaller for the Laplace distribution, but the posterior distributions of the estimated trends are similar.
Almost identical posterior profiles are obtained for $b$=0.05, 0.10 and 0.20 in the left plots.
These results are also very similar to those obtained using the Pearson family of distributions.

\subsection{Mixture Distributions}

Given two probability distributions $f(x)$ and $g(x)$, the mixture distribution is defined by $h(x)=\alpha f(x)+(1-\alpha )g(x)$, where $\alpha$ is the mixture weight. Here, we consider the cases where $f(x)$ and $g(x)$ are either Gaussian distribution, uniform distribution and $\delta$-function.
Although the $\delta$-function is different from an ordinary probability distribution, it is equivalent to not adding system noise, so implementation with a non-Gaussian filter is straightforward.

\begin{figure}[tbp]
\begin{center}
\includegraphics[width=130mm,angle=0,clip=]{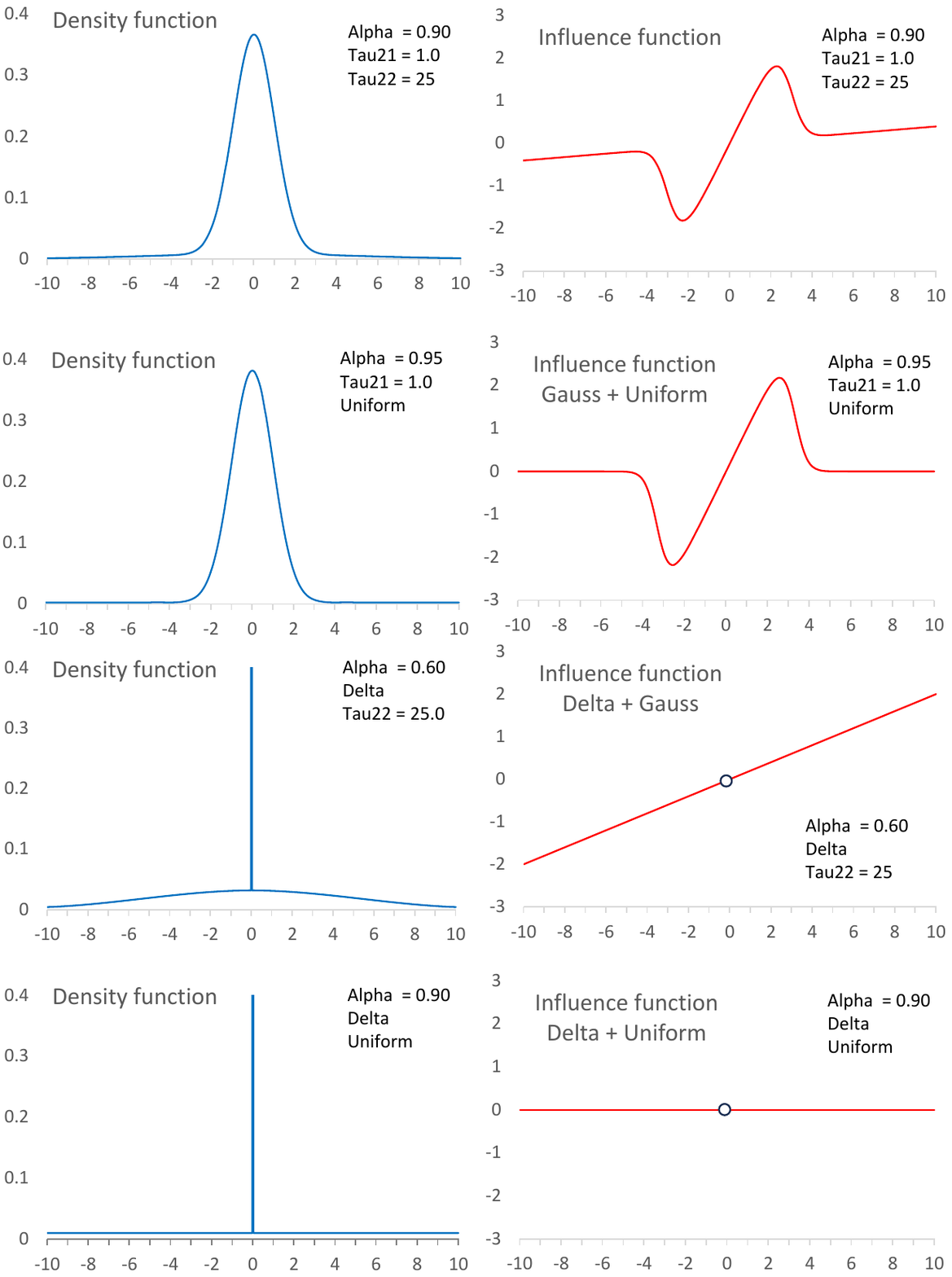}
\caption{Density functions and influence functions of various mixture models.
The mixture distributions of two Gaussian distributions, Gaussian and uniform distributions, 
Gaussian distribution and $\delta$-function, and $\delta$ and uniform distributions, from top to bottom.}
\label{figure:Influence-function_Mixture}
\end{center}
\end{figure}

Figure \ref{figure:Influence-function_Mixture} shows the density functions and influence functions of the mixing distributions. From top to bottom, four cases are shown: mixture of a Gaussian distribution with small variance and a Gaussian distribution with large variance, mixture of a Gaussian distribution and a uniform distribution, mixture of a $\delta$-function and a Gaussian distribution, and mixture of a $\delta$-function and a uniform distribution.
The two cases below are not differentiable at the origin ($x$=0) and should be considered as conceptual diagrams.
In the case of a mixture of two Gaussian distributions, the influence function is a smooth connection of two straight lines with different slopes.

\begin{table}[tbp]
\caption{Trend model with various noise distributions. Gaussian, Laplace, Cauchy, Pearson and generalized Laplace distributions and mixture distributions of two Gaussian distributions, Gaussian and uniform distributions, delta and uniform distributions and delta and Gaussian distributions.}
\label{Tab_summary_of_mixture_distribution_models}
\tabcolsep=2mm
\begin{center}\begin{tabular}{ccccccc}
Distribution  & $\sigma^2$ & $\tau^2$ & $b$ or $\alpha$   & log-LK & $k$ & AIC \\
\hline
Gaussian & 1.0439 & $1.249\times 10^{-2}$ &  ---   & $-748.5156$ &  2 & 1,501.031 \\
Laplace  & 1.0359 & $1.233\times 10^1\,\,\,$       &  ---   & $-747.9692$ &  2 & 1,499.938 \\
Cauchy   & 1.0227 & $3.454\times 10^{-5}$ &  ---   & $-742.1673$ &  2 & {\color{blue}1,488.335} \\
Pearson  & 1.0241 & $7.411\times 10^{-8}$ & 0.7823 & $-741.8784$ &  3 & 1,489.757 \\
G-Laplace& 1.0255 & $1.665\times 10^1\,\,\,$         & 0.0913 & $-741.7879$ &  3 & 1,489.576 \\
\hline
G(0,$\tau^2$)+G(0,4)  & 1.0265 & $3.086\times 10^{-4}$ & 0.9905 & $-741.7726$ & 3 & 1,489.545 \\
G(0,$\tau^2$)+U[-4;4] & 1.0264 & $3.965\times 10^{-4}$ & 0.9923 & $-743.0081$ & 3 & 1,492.016 \\
$\delta$(0)+U[-4;4]   & 1.0338 &       ---             & 0.9998 & $-743.2806$ & 2 & 1,490.561 \\
$\delta$(0)+G(0,4)    & 1.0315 &       ---             & 0.9900 & $-741.9420$ & 2 & {\color{red}1,487.884} \\
\hline
\end{tabular}\end{center}
\end{table}

Table \ref{Tab_summary_of_mixture_distribution_models} compares mixture noise models and representative distribution models. The number of parameters for the Gaussian, Laplace and Cauchy distribution models is $k$=2, while for the Pearson and generalised Laplace distributions the shape parameter $b$ is also used for maximum likelihood estimates, so the number of parameters is 3.

The table shows that the AIC of the mixture distribution model with $\delta$-function and Gaussian distribution is minimal. However, the AIC's of the Cauchy distribution model, the Pearson distribution model, the generalised Laplace distribution model and the mixture of two Gaussian distribution models differ from the minimum AIC by less than two, so it should be considered that their superiority is not so clear.

\begin{figure}[tbp]
\begin{center}
\includegraphics[width=75mm,angle=0,clip=]{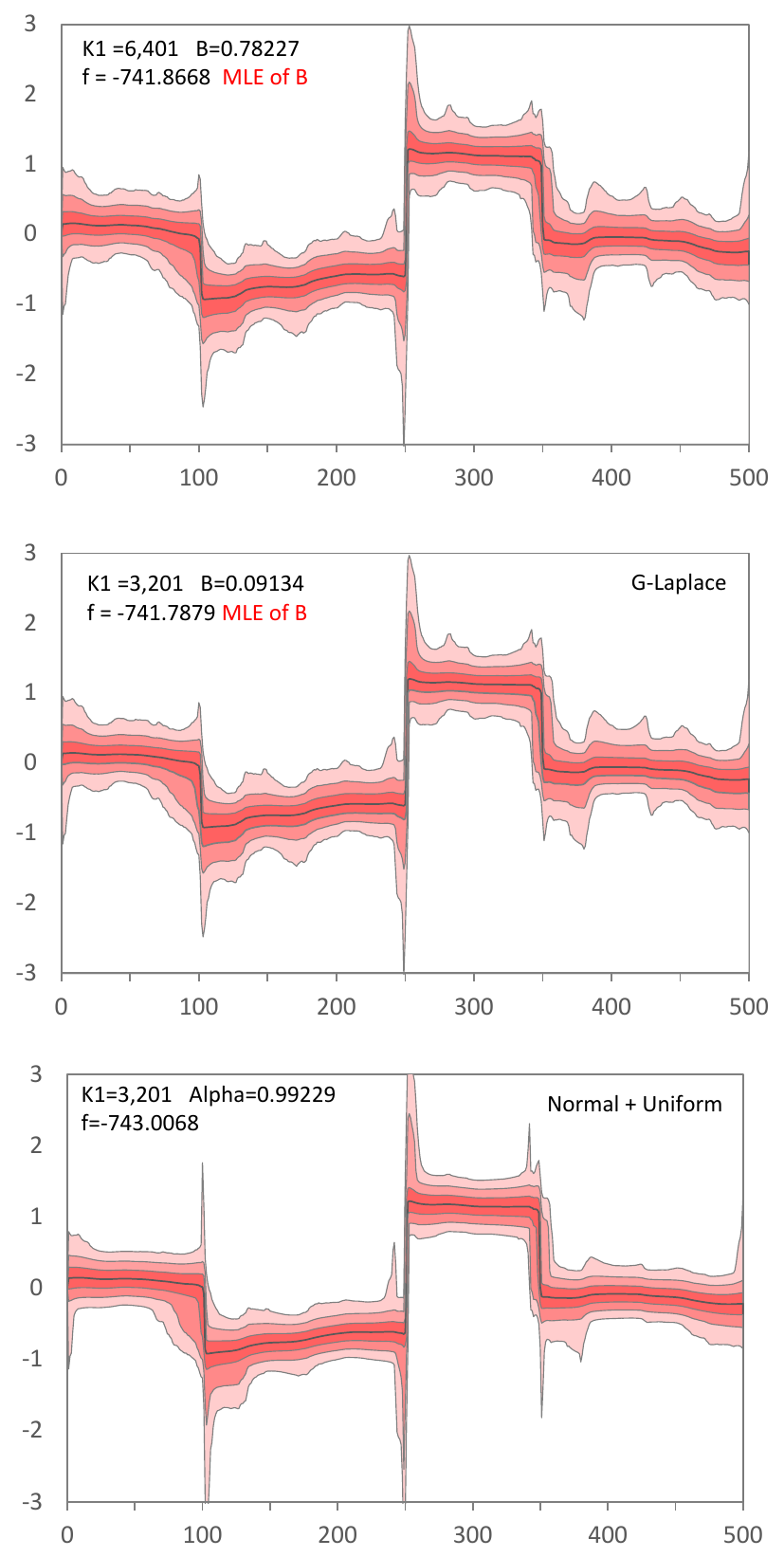}
\includegraphics[width=75mm,angle=0,clip=]{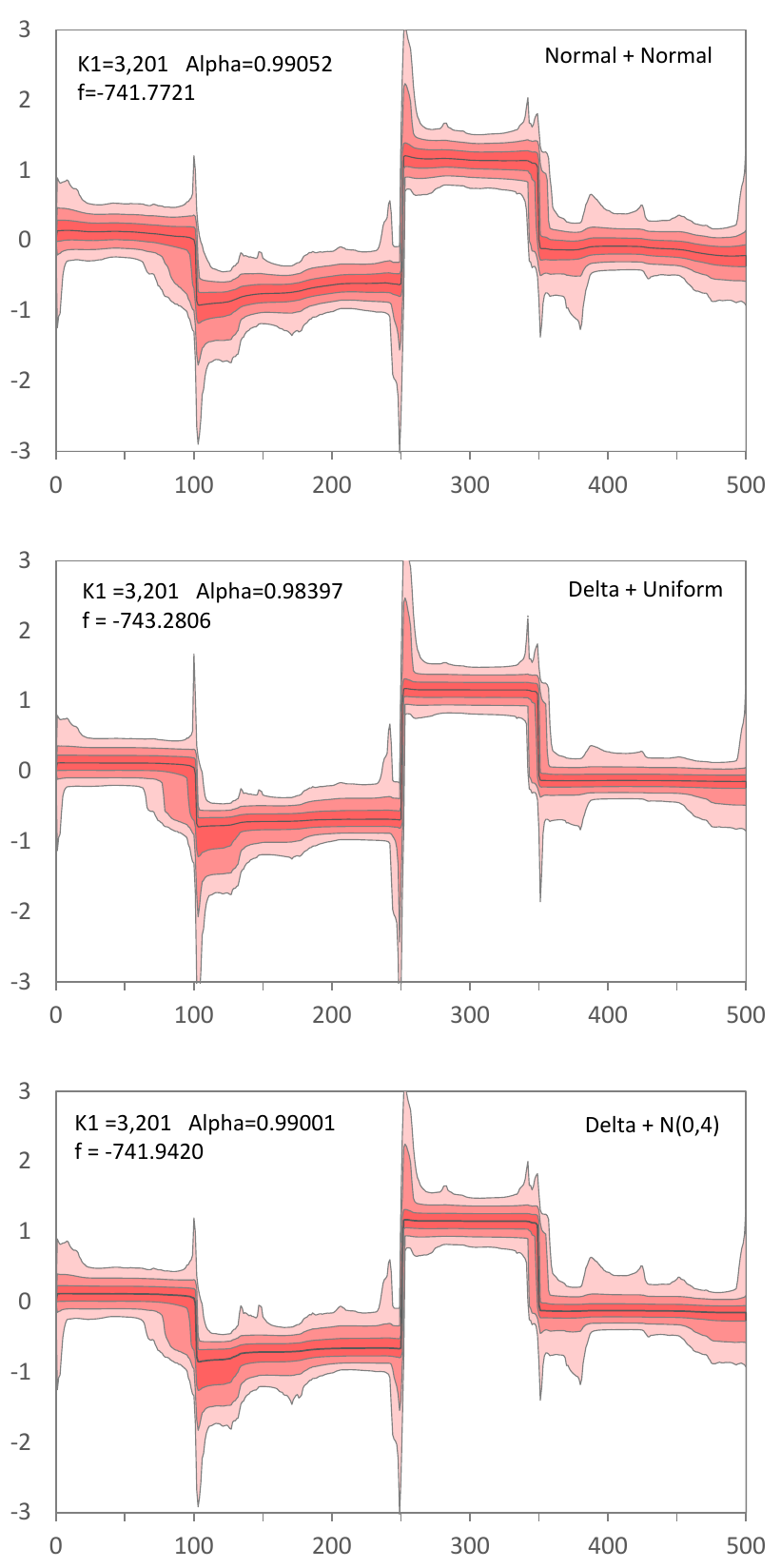}
\caption{Trend estimation by self-organizing state-space model implemented by non-Gaussian filter and smoother with mixture models. From upper left to lower right: the Pearson type VII distribution with MLE of $b$, generalized Laplace distribution with MLE of $b$, mixture of Gaussian and uniform distributions, mixture of two Gaussian distributions, mixture of delta (point mass) and uniform distribution, and mixture of delta and Gaussian distributions.}
\label{figure:trend_Mixture}
\end{center}
\end{figure}

Figure \ref{figure:trend_Mixture} shows the trend estimation results from the Pearson distribution model, the generalised Laplace distribution model and the four mixture distribution models. Three jumps are clearly detected in all models, with very smooth trends at the other locations.

However, a careful look reveals that the trend obtained by the four mixture distribution models is smoother. In particular, when using the mixture distribution model with $\delta$ and Gaussian or uniform distribution, the obtained trend is almost constant except at the jump points.
A mixed distribution model with a $\delta$-function is considered appropriate when the trend is usually constant and changes rarely and abruptly.



\section{Concluding Remarks}

This study examined various probability distributions for modeling time series data with sudden structural changes, such as abrupt shifts in trends, and reported the results of their application to test data.

The Pearson type VII family of distributions, which can capture a range of shapes from Gaussian to heavily-tailed distributions, was found to be particularly versatile. By using the maximum likelihood method, the shape parameter $b$ can be flexibly adjusted to suit specific situations.

Key findings include the following:

\begin{enumerate}
\item The generalized Laplace distribution, despite being non-differentiable at the origin -- potentially complicating parameter estimation -- produced results nearly identical to the Pearson distribution model in practice. In some cases, it offered improved performance based on likelihood and AIC criteria.
\item In mixture distribution models, the highest likelihood was observed for two configurations: a mixture of two Gaussian distributions and a combination of a $\delta$-function with a Gaussian distribution featuring a large variance. However, the latter model, having one fewer parameter, was favored from an AIC perspective. Notably, the inclusion of a $\delta$-function appears particularly effective in scenarios where trends are generally stable, with infrequent but significant jumps.
\item Estimation using mixture distribution models tended to be relatively unstable. In contrast, the Pearson VII or extended Laplace distributions provide more robust alternatives for general cases.
\end{enumerate}

Overall, these findings highlight the importance of selecting appropriate models based on the specific characteristics of the time series data and the goals of the analysis.


\vspace{10mm}
\begin{thebibliography}{3}


\bibitem{RefB}
Anderson, B.D.O., and Moore, J.B. (1979).
{\em Optimal Filtering},
New Jersey, Prentice-Hall.



\bibitem{RefJ}
Doucet, A., Godsill, S. and Andrieu, C. (2000).
On sequential Monte Carlo ampling methods for Bayesian filtering,
{\it Statistics and Computing}, {\bf 10}, 197--208.


\bibitem{RefB}
Doucet, A.,  de Freitas, N., and Gordon, N., (2001).
    {\it Sequential Monte Carlo Methods in Practice}. 
    Springer-Verlag, New York.

\bibitem{DK 2012}
Durbin, J., and Koopman, S. J., (2012). 
\textit{Time Series Analysis by State Space Methods}, Vol.38. 
Oxford Statistical Science, Croydon, UK.


\bibitem{RefJ}
Gordon, N.~J., Salmond, D.~J., and Smith, A.~F.~M., (1993). 
    Novel approach to nonlinear/non-Gaussian Bayesian state estimation, 
    {\it IEE Proceedings--F}, {\bf 140}, 107--113.



\bibitem{HRRS 1986}
Hampel, F.R., Ronchetti, E.M.,Rousseeuw, P.J. and Stahel, W.A. (1986).
\textit{Robust Statistics: The Approach Based on Influence Functions}, 
Wiley Series in Probability and Statistics, 
John Wiley \& Sons, New York.

\bibitem{RefJ}
Harrison, P.J. and Stevens, C.F. (1976),  Bayesian Forecasting (with discussion), {\it Journal of the Royal Statistical Society}, Series B, Vol. 34, 1-41.




\bibitem{RefJ}
 Kitagawa, G. (1987),   Non-Gaussian State Space Modeling of
Nonstationary Time Series, {\em Journal of American Statistical Association}, Vol.76, No.400, 1032-1064.

\bibitem{RefJ}
 Kitagawa, G. (1988), Numerical Approach to Non-Gaussian Smoothing and its Applications,
{\it Computing Science and Statistics; Proceedings of the 20th Symposium on the Interface,} eds. E.J. Wegman, D.T. Gantz and J.J. Miller, 379-388.




\bibitem{RefJ}
Kitagawa, G., (1996).  Monte Carlo filter and smoother for non-Gaussian nonlinear state space model,
   {\it Journal of Computational and Graphical Statistics}, {\bf 5}, 1--25.

\bibitem{RefJ}
Kitagawa, G., (2020). {\it Introduction to Time Series Modeling with Applications in R}, 
Chapman \& Hall/CRC Press, New York.


\bibitem{RefJ}
Kitagawa, G. and Gersch, W.(1984). A Smoothness
Priors-State Space Approach
to the Modeling of Time Series with Trend and Seasonality,
{\it Journal of the American Statistical Association}, 79, No.386, 378-389.

\bibitem{RefB}
Kitagawa, G., and Gersch, W., (1996). 
  {\it Smoothness Priors Analysis of Time Series}, 
  Springer-Verlag, New York.


\bibitem{KKP 2001}
Kotz, S., Kozubowski, T. and Podg\'{o}rski, K. (2001).
\textit{The Laplace Distribution and Generalizations; A Revisit with New Applications},\\
https://www.researchgate.net/publication/258697410







\bibitem{SS 2000}
Shumway, R.H., and Stoffer, D.S., (2000).
\textit{Time Series Analysis and Its Applications}, 
Springer Texts in Statistics, Springer, New York.



\bibitem{RefB}
 West, M. and Harrison, J. (1989).  {\em Bayesian Forecasting and Dynamic Models}
, Springer Series in statistics, Springer-Verlag, New York.
 


\end{thebibliography}
\end{document}